\documentclass[showpacs]{revtex4}
\usepackage{graphicx,psfrag,amsmath,amssymb,amsfonts,latexsym,color,dcolumn,
graphpap}
\begin{document}
\title{Derivative expansion for the boundary interaction terms in the Casimir
effect: generalized $\delta$-potentials} 
\author{C. D. Fosco$^a$}
\author{F. C. Lombardo$^b$ }
\author{F. D. Mazzitelli$^b$ }
\affiliation{$^a$Centro At\'omico Bariloche and Instituto Balseiro,
 Comisi\'on Nacional de Energ\'\i a At\'omica, \\
R8402AGP Bariloche, Argentina}
\affiliation{$^b$Departamento de
F\'\i sica {\it Juan Jos\'e Giambiagi}, FCEyN UBA, Facultad de
Ciencias Exactas y Naturales, Ciudad Universitaria, Pabell\' on I,
1428 Buenos Aires, Argentina.}
\date{today}
\begin{abstract}
\noindent We calculate the Casimir energy for scalar fields in interaction
with finite-width mirrors, described by nonlocal interaction terms. These
terms, which include quantum effects due to the matter fields inside the
mirrors, are approximated by means of a local expansion procedure.  As a
result of this expansion, an effective theory for the vacuum field emerges,
which can be written in terms of generalized $\delta$-potentials.  We
compute explicitly the Casimir energy for these potentials and show that,
for some particular cases, it is possible to reinterpret them as imposing
imperfect Dirichlet boundary conditions.
\end{abstract}
\pacs{03.70.+k; 11.10.Gh; 42.50.Pv; 03.65.Db}
\maketitle
\section{Introduction}\label{sec:intro}

Casimir forces are a striking manifestation of the zero-point energy of the
electromagnetic field in the presence of `mirrors' endowed with quite
general electromagnetic properties~\cite{reviews}. In many calculations of
the Casimir energies and forces, the presence of the mirrors is modeled by
appropriate boundary conditions on the interfaces of the different media,
that include  macroscopic parameters such as their electric permitivity,
magnetic permeability, conductivity, etc.  A first-principles calculation
of the Casimir energy should consider the microscopic degrees of freedom
associated to the mirrors. This could shed light on some interesting open
questions, the role of dissipation on the Casimir energy being, perhaps,
the most important among them.

In a previous paper~\cite{Fosco:2008td}, we considered the Casimir effect
for scalar and gauge fields interacting with dynamical matter on thin mirrors
(see also Ref.\cite{graphene} for a concrete model realization). 
More recently, one of us considered the generalization to the case of
finite-width mirrors~\cite{Fosco:2009cw}. The interaction between the
vacuum scalar field and the mirrors' degrees of freedom gives rise, in
general, to a nonlocal effective action in terms of which the Casimir
energy may be calculated~\cite{extra}. Moreover, under certain
circumstances, it is possible to find a formal expression for the Casimir
energy in terms of the parameters that define the nonlocal
kernel~\cite{Fosco:2009cw}.  In this paper, we will present an application
of the previously developed formalism for the Casimir effect with nonlocal
boundary interaction terms, to situations where those nonlocal terms may be
expanded in a series of local ones.  In other words, we will perform a
derivative expansion of the nonlocal effective action.  One should expect
on physical grounds that, in many relevant cases, such a local description
of the mirrors must be reliable. We show here how one can indeed find such
an expansion, and then we shall apply it to derive approximate 
expressions for the Casimir energy. 

The structure of this paper is the following. In Section~\ref{sec:scalar},
we derive the derivative expansion for the nonlocal effective action, which will be
written in terms of a set of generalized $\delta$-potentials, i.e. terms
proportional to Dirac's $\delta$-function and its derivatives. We will
illustrate, in  concrete examples, how the coupling between the vacuum
field and the microscopic degrees of freedom, together with the boundary
conditions that confine the microscopic degrees of freedom inside the
mirrors, do determine the different coefficients in the derivative
expansion. In Section \ref{sec:deltapot} we compute the Casimir energy for the resulting
generalized $\delta$-potentials.  Section \ref{sec:disc} contains our final remarks.

\section{Derivative expansion of the nonlocal effective action}\label{sec:scalar}

Let us  consider a real scalar field $\varphi$ in the presence of two flat mirrors 
of width $\epsilon$ centered at $x_d=0,\, a$. This scalar field interacts
with  the microscopic degrees of freedom inside the mirrors, which in the
specific examples below will be described by a second scalar field $\chi$.
After integrating the microscopic degrees of freedom of the mirrors, the
effective action for the scalar field $\varphi$ will be of the form
\begin{equation}
 S(\varphi) =  \frac{1}{2} \int d^{d} x_\parallel dx_d
 \left(\partial \varphi\right)^2 + S_I^{(0)}(\varphi) + S_I^{(a)}(\varphi),
\end{equation}
where $S_I^{(0)}$ and $S_I^{(a)}$ are concentrated on the positions of each mirror.
On general grounds we expect these interaction terms to be nonlocal, i.e.,
\begin{equation}\label{eq:nonlocal}
S_I^{(0)}(\varphi) \;=\; \frac{1}{2} \, \int_{-\infty}^{+\infty} dx_d 
\int_{-\infty}^{+\infty}dx'_d \int \frac{d^d k_\parallel}{(2\pi)^d} \;
{\widetilde\varphi}^*(k_\parallel,x_d) 
{\widetilde V}_\epsilon(k_\parallel;x_d,x'_d) 
{\widetilde\varphi}(k_\parallel,x'_d) \;,
\end{equation}
and a similar expression for $S_I^{(a)}$. Here $x_\parallel$ denotes the
time ($x_0$) as well as the $d-1$ spatial coordinates parallel to the
mirror (which we shall denote by $\mathbf{x_\parallel}$). We have assumed
translational invariance in the coordinates $x_\parallel$, and therefore
it is useful to write the effective action in terms of the Fourier transform  of the field
in these coordinates,  $\widetilde\varphi$, with the obvious notation  $k_\parallel$ for  the argument of this function. 

The
nonlocal kernel ${\widetilde V}_\epsilon$ may be expanded as follows \cite{Fosco:2009cw}
\begin{equation}\label{eq:vnl}
{\widetilde V}_\epsilon(k_\parallel;x_d,x'_d) \;=\; \sum_{m,n}
\psi_m^{(\epsilon)}(x_d) \, C_{mn}(k_\parallel,\epsilon)
\psi_n^{(\epsilon)*}(x'_d) \;,
\end{equation}
and the functions $\psi_n^{(\epsilon)}(x_d)$ depend essentially on the
nature of the boundary conditions for the microscopic fields (i.e., those
living inside the mirrors) while the coefficients,
$C_{mn}(k_\parallel,\epsilon)$, are obtained by taking into account the
(kinematic and dynamical) properties of those fields.  Eq. (\ref{eq:nonlocal}) results from the assumption that, after integrating the
microscopic fields, the most relevant term in the effective action is
quadratic in the scalar field; in other words, we are assuming, as usual,
that the media can be described by linear response theory. The particular case
in which the interaction between the thick mirrors and the vacuum field is approximated by a local
effective action (i.e. ${\widetilde V}_\epsilon(k_\parallel;x_d,x'_d)=f(k_\parallel,x_d)\delta(x_d-x'_d)$) has
been considered in Ref.\cite{prd08}.

As we will see, the nonlocal effects can be evaluated perturbatively by expanding the kernel
${\widetilde V}_\epsilon$ in powers of the $\delta$-function and its derivatives:
\begin{equation}\label{eq:vnldelta}
{\widetilde V}_\epsilon(k_\parallel;x_d,x'_d) \;=\;\tilde\mu_0(k_{\parallel})\delta(x_d)\delta(x'_d)
+\tilde\mu_1(k_{\parallel})
\left( \delta(x_d)\delta ' (x'_d) +\delta ' (x_d)\delta  (x'_d)\right)
+ \tilde\mu_2(k_{\parallel})\delta ' (x_d)\delta '(x'_d) + ...
\end{equation}
where $\tilde\mu_i(k_{\parallel})$ depend on the microscopic fields and their interaction with the
vacuum field.

We start our derivation of the expansion with the study of a simple example,
namely, the case in which the microscopic field $\chi$ is also a real
scalar, endowed with a quadratic action, and linearly coupled to $\varphi$. As already
mentioned, we 
denote by $\epsilon$ the width of the mirror, which fills the region 
$-\epsilon/2\leq x_d\leq\epsilon/2$.
Then, as shown in~\cite{Fosco:2009cw}, the coefficients
$C_{mn}(k_\parallel,\epsilon)$ adopt the  diagonal form:
\begin{equation}\label{eq:vnl2}
C_{mn}(k_\parallel,\epsilon) \;=\; \frac{ g^2}{ \xi_n^2 +
k_\parallel^2 +m^2} \, \delta_{mn}\;, 
\end{equation}
where $m$ is the mass of the microscopic field, $\xi_n^2$ ($\xi_n \in
{\mathbb R}$) denote the eigenvalues of $(-\partial_d^2)$ corresponding to
the eigenvectors $\psi_n^{(\epsilon)}(x_d)$, and $g$ is the coupling
constant between $\varphi$ and $\chi$. 

The precise form of those eigenvalues and eigenvectors depends of course on
the boundary conditions for the microscopic field. Indeed, for the case of
Dirichlet boundary conditions, we have the eigenfunctions:
\begin{equation}\label{eq:psid}
 \psi_n^{(\epsilon)}(x_d) \;=\; \sqrt{\frac{2}{\epsilon}} \,\times \, 
\left\{ 
\begin{array}{lcll}
\sin(\frac{n\pi x_d}{\epsilon}) & {\rm if}& \; n= 2 k, & (k = 1, 2,
\ldots)\\ 
\cos(\frac{n \pi x_d}{\epsilon}) & {\rm if}& \; n= 2 k + 1, & (k = 0, 1,
\ldots )\;,
\end{array}
\right.
\end{equation}
while in the Neumann case, we have instead
\begin{equation}\label{eq:psin}
 \psi_n^{(\epsilon)}(x_d) \;=\; \frac{1}{\sqrt{\epsilon}}\times \, 
\left\{ 
\begin{array}{ccll}
1 & {\rm if} &  n = 0 , & \\
\sqrt{2} \, \sin(\frac{n\pi x_d}{\epsilon}) & {\rm if}& \;
n= 2 k + 1, & (k = 0, 1, \ldots)\\ 
\sqrt{2} \, \cos(\frac{n \pi x_d}{\epsilon}) & {\rm if}&
\; n= 2 k , & (k = 1, 2, \ldots )\;.
\end{array}
\right.
\end{equation}
The eigenvalues are then, in both cases, given by the expression: 
\mbox{$\xi_n^2 = \big(\frac{\pi n}{\epsilon}\big)^2$}, where $n =
1,2,\ldots$ in the Dirichlet case, while for Neumann boundary conditions:
$n = 0,1,2,\ldots$ 

The essential difference is thus the existence or not of a zero mode, which
is present only in the Neumann case. One should expect this difference to
manifest itself when one tries to perform a local approximation for the
nonlocal interaction term, under the assumption that $\epsilon \to 0$.
Indeed, note that, in such a case, the zero mode is multiplied by the
$\epsilon$-independent coefficient $C_{00}$, while all the other $C_{mn}$
coefficients are relatively suppressed in such a limit. To make this
statement more precise,  let us perform an expansion of the interaction
term, assuming that $\epsilon$ is small (this shall be made more clear
below, after introducing the other length scale to compare it with).  

We shall assume, in what follows, Neumann boundary conditions for the
microscopic field. Writing $S_I^{(0)}$ more explicitly:
\begin{equation}\label{eq:nonlocal1}
S_I^{(0)}(\varphi) \;=\; \frac{1}{2} \, g^2 \,  \int \frac{d^d k_\parallel}{(2\pi)^d} \;
\sum_{m,n=0} \, \langle {\widetilde\varphi} |\psi_n^{(\epsilon)} \rangle
\,\frac{1}{ \xi_n^2 + k_\parallel^2 + m^2} \, \langle \psi_n^{(\epsilon)}
|{\widetilde\varphi} \rangle \;,
\end{equation}
where we used the notation \mbox{$\langle f | g \rangle \equiv
\int_{-\frac{\epsilon}{2}}^{+\frac{\epsilon}{2}} dx_d \, f^*(x_d) \,
g(x_d)$}.

Let us first consider the leading term in a small-$\epsilon$ expansion,
obtained by keeping only the zero mode contribution. This may be written as
follows:
\begin{equation}\label{eq:local}
S_I^{(0)}(\varphi) \,\approx \, \frac{1}{2} \, g^2 \epsilon \,  \int \frac{d^d
k_\parallel}{(2\pi)^d} \, \frac{1}{k_\parallel^2+m^2} \; 
\int_{-\infty}^{+\infty} dx_d \,\delta_\epsilon (x_d)  {\widetilde\varphi}^*(k_\parallel,x_d) 
\int_{-\infty}^{+\infty}  dx'_d \, \delta_\epsilon(x'_d) {\widetilde\varphi}(k_\parallel,x'_d) 
\end{equation}
where we introduced $\delta_\epsilon (x_d) \equiv \theta
\big(\frac{\epsilon}{2} - |x_d|\big)/\epsilon$, which works as an
approximant of Dirac's $\delta$-function. Then we see that this leading
term may be regarded as a local $\delta$-function term, with a
momentum-dependent strength, and affected by a coefficient $g^2 \epsilon$.
It is convenient to introduce the product $g^2 \epsilon \equiv \lambda$,
since that is the constant that determines the strength of the boundary
interaction term:
\begin{equation}\label{eq:locallimit}
S_I^{(0)}(\varphi) \,\to \, \frac{1}{2} \, \lambda \,  \int \frac{d^d
k_\parallel}{(2\pi)^d}\int_{-\infty}^{+\infty} dx_d  \,
\frac{1}{k_\parallel^2+m^2} \; 
\delta(x_d) \,{\widetilde\varphi}^*(k_\parallel,x_d)  {\widetilde\varphi}(k_\parallel,x_d) \;.
\end{equation}

The following terms in the expansion are obtained by expanding  
the overlaps $\langle \psi_n^{(\epsilon)} |\widetilde{\varphi} \rangle$ 
for small $\epsilon$, by using a Taylor expansion
for the vacuum field. Up to the second order in derivatives, we find:
\begin{equation}\label{eq:nonlocal3}
\langle \psi_n^{(\epsilon)} |\widetilde{\varphi} \rangle \;=\;
\sqrt{\epsilon} \times\, \left\{
\begin{array}{cll}
{\widetilde\varphi}(k_\parallel,0) & {\rm if} \; n=0 & \; \\
\sqrt{2} \,\frac{\epsilon}{\pi^2} \, \frac{2}{(2k+1)^2} \, (-1)^k
\;\big[\partial_d{\widetilde\varphi}(k_\parallel,x_d)\big]_{x_d=0} & {\rm
if} \; n = 2 k + 1 & (k =0,1,\ldots) \\ 
\sqrt{2} \big(\frac{\epsilon}{\pi}\big)^2 \, \frac{1}{(2k)^2} \, (-1)^k
\;\big[\partial^2_d{\widetilde\varphi}(k_\parallel,x_d)\big]_{x_d=0} & {\rm
if} \; n = 2 k & (k =1,2,\ldots ) 
\end{array}
\right. \;.
\end{equation} 
We see that the assumption that one could use to justify the expansion {\em
a posteriori\/} is that the field inside the mirror should not change
appreciably inside the mirror, more precisely, the length scale of the
spatial variation should be much larger than $\epsilon$. This means
that higher powers of $\epsilon$ will be attached to higher derivatives of
the field. 

Equipped with the expansion (\ref{eq:nonlocal3}), we obtain the
corresponding expansion of $S_I^{(0)}$ up to the second order in
derivatives: 
\begin{equation}\label{eq:exp1}
S_I^{(0)} \;=\; S_{I,0}^{(0)} \,+\,S_{I,2}^{(0)}\,+\, \ldots 
\end{equation}
where  $S_{I,0}^{(0)}$ coincides with Eq.(\ref{eq:local}), while:
\begin{equation}\label{eq.second}
S_{I,2}^{(0)}=  \frac{1}{2} \, g^2 \, 8 \, \epsilon
\,(\frac{\epsilon}{\pi})^2 \, \int \frac{d^d k_\parallel}{(2\pi)^d} \;
\big| {\widetilde \varphi}'(k_\parallel,0)\big|^2 \,
\sum_{l=0}^{\infty} \frac{1}{(2 l + 1)^2 \Big[(2 l +1)^2
(\frac{\pi}{\epsilon})^2 + k_\parallel^2 + m^2\Big]} \;.
\end{equation}
Here the prime denotes derivative with respect to $x_d$. 
Performing the sum of the series, we obtain
\begin{equation}\label{eq.second1}
S_{I,2}^{(0)}=  \frac{1}{2} \, g^2 \, 8 \, \epsilon^3 \, 
\int \frac{d^d k_\parallel}{(2\pi)^d} \;
\big| {\widetilde \varphi}'(k_\parallel,0)\big|^2 \,
\frac{1}{\big[\omega(k_\parallel)\big]^2} \, \Big\{ \frac{1}{8} - \frac{1}{4 \epsilon
\, \omega(k_\parallel)} \, \tanh\big[ \frac{\epsilon \,
\omega(k_\parallel)}{2}\big]\Big\} \;,
\end{equation}
where $\omega(k_\parallel) \equiv \sqrt{k_\parallel^2 + m^2}$.
We may again write this term in a similar fashion to (\ref{eq:local}):
\begin{equation} \label{eq:local2}
S_{I,2}^{(0)}(\varphi) \,\approx \, \frac{1}{2} \, \lambda\,  \int \frac{d^d
k_\parallel}{(2\pi)^d} \, f(k_\parallel,\epsilon) \,
\frac{1}{\big[\omega(k_\parallel)\big]^2} \; 
\int_{-\infty}^{+\infty} dx_d \,\delta'(x_d)  {\widetilde\varphi}^*(k_\parallel,x_d) 
\int_{-\infty}^{+\infty}  dx'_d \, \delta'(x'_d) {\widetilde\varphi}(k_\parallel,x'_d) 
\end{equation}
with
\begin{equation}\label{eq.second2}
f(k_\parallel,\epsilon) \;=\; \epsilon^2 \, 
\Big\{ 1- \frac{2}{\epsilon \, \omega(k_\parallel)} \, 
\tanh\big[ \frac{\epsilon \omega(k_\parallel)}{2}\big] \Big\} \;.
\end{equation}
We will now summarize the result for the expansion of
the nonlocal term, presenting it in a way which shall be useful in the
derivation of the Casimir energy. Thus, we encode the results as follows:
the expansion for the interaction action (\ref{eq:nonlocal}) may be
interpreted as an expansion for the nonlocal potential, so that:
\begin{equation}
{\widetilde V}_\epsilon(k_\parallel;x_d,x'_d) \;=\; 
{\widetilde V}_{\epsilon,0}(k_\parallel;x_d,x'_d) \,+\, {\widetilde
V}_{\epsilon,2}(k_\parallel;x_d,x'_d) \,+\,\ldots 
\label{expansionV}\end{equation}
where:
\begin{eqnarray}
{\widetilde V}_{\epsilon,0}(k_\parallel;x_d,x'_d) &=&
\tilde\mu_0(k_\parallel,\epsilon) \, \delta(x_d) \, \delta(x'_d) \nonumber\\
{\widetilde V}_{\epsilon,2}(k_\parallel;x_d,x'_d) &=&
\tilde\mu_2(k_\parallel,\epsilon) \, \delta'(x_d) \, \delta'(x'_d) 
\label{expansionV0y2}\end{eqnarray}
with $\tilde\mu_0(k_\parallel,\epsilon) =
\frac{\lambda}{[\omega(k_\parallel)]^2}$ and 
$\tilde\mu_2(k_\parallel,\epsilon) = \frac{\lambda f(k_\parallel,\epsilon)}{[\omega(k_\parallel)]^2}$. 

It is worth noting that the derivative expansion of the effective action
for the specific example considered so far does not include terms
containing  only one derivative of the scalar field (i.e. terms
proportional to $\tilde\mu_1(k_\parallel)$, see Eq.(\ref{eq:vnldelta})).  This is
due to the fact that the eigenfunctions in Eqs.(\ref{eq:psid}) and
(\ref{eq:psin}) have a definite parity in the interval $-\epsilon/2\leq
x_d\leq\epsilon/2$. Therefore, we expect such terms to show up only if one
considered `non-symmetric' mirrors in which the boundary conditions that
confine the microscopic fields are different on both interfaces
$x_d=\pm\epsilon/2$, for example $\chi(-\epsilon/2)=0,
\partial_d\chi(\epsilon/2)=0$.

Up to here, we considered a microscopic field $\chi$ linearly coupled to
$\varphi$. The expansions presented in Eqs.(\ref{expansionV}) and
(\ref{expansionV0y2}) remain valid when the microscopic $\chi$-field is
nonlinearly coupled to $\varphi$.  To illustrate this fact, we will now
consider a different case, i.e.  we shall deal with a $g \chi^2 \varphi$
coupling term for $d=3$. In terms of the eigenfunctions $\psi_n^{(\epsilon)}$, the
second order term in the expansion of the action $S_I(\varphi)$ is given
by:
\begin{equation}
S^{(0)}_I \approx  g^2 \int \frac{d^3k_\parallel}{(2\pi)^3} \left\vert \tilde\varphi(k_\parallel,0)
\right\vert^2 \int \frac{d^3p_\parallel}{(2\pi)^3}
\int_{-\infty}^{+\infty}dx_d\int_{-\infty}^{+\infty}dx_d' \sum_{n,m}
\frac{\psi_n^{(\epsilon)}(x_d) \psi_n^{(\epsilon)}(x_d')
\psi_m^{(\epsilon)}(x_d) \psi_m^{(\epsilon)}(x_d')}{\left(\xi_n^2 + m^2 +
p_\parallel^2\right) \left(\xi_m^2 + m^2 + (p_\parallel +
k_\parallel)^2\right)} \;.
\end{equation}
The integrations over normal coordinates ($x_d$ and $x_d'$) can be trivially performed by means of the 
orthogonality conditions of the eigenfunctions $\psi_n^{(\epsilon)}$, obtaining
\begin{equation}
S^{(0)}_I \approx  g^2 \int \frac{d^3k_\parallel}{(2\pi)^3} \left\vert
\tilde\varphi(k_\parallel,0)\right\vert^2 \sum_{n} \int
\frac{d^3p_\parallel}{(2\pi)^3} \frac{1}{\left(\xi_n^2 + m^2 +
p_\parallel^2\right) \left(\xi_n^2 + m^2 + (p_\parallel +
k_\parallel)^2\right)}\;.
\end{equation}

After integrating over $p_\parallel$, one sees that the interaction action
becomes
\begin{equation}
S^{(0)}_I \approx  \frac{g^2}{4\pi} \int \frac{d^3k_\parallel}{(2\pi)^3} \left\vert 
\tilde\varphi(k_\parallel,0)\right\vert^2 \sum_{n} \frac{1}{k_\parallel} 
\arctan{\left\{\frac{k_\parallel}{2\sqrt{\xi_n^2 + m^2}}\right\}},
\end{equation}
whence one can read the coefficient $\tilde\mu_0(k_\parallel, \epsilon)$ in Eq.(\ref{expansionV0y2}), 
\begin{equation}
 \tilde\mu_0(k_\parallel, \epsilon) = \frac{g^2}{4\pi} 
\frac{1}{k_\parallel}\sum_{n}\arctan{\left\{\frac{\epsilon k_\parallel}{2\sqrt{n^2\pi^2 + \epsilon^2 m^2}}\right\}},
\label{mu0}\end{equation}
where we have used that $\xi_n^2 = \left(\frac{n \pi}{\epsilon}\right)^2$. 

The sum in last equation runs from $0$ for Neumann boundary conditions, and
from $1$ for Dirichlet ones.  Eq.(\ref{mu0}) is divergent for large
$n$, therefore we introduce a renormalization term, in order to obtain a
finite coefficient. For a massless field with Dirichlet boundary conditions
we obtain:
\begin{equation}
 \tilde\mu_0(k_\parallel, \epsilon) = {\bar\mu}_0 + \frac{g^2}{4\pi} 
\sum_{n\geq 1}\left[\frac{1}{k_\parallel}\arctan{\big(\frac{\epsilon k_\parallel}{2 n \pi}\big)} - \frac{\epsilon}{2 n \pi}\right],
\end{equation}
where ${\bar\mu}_0$ is a renormalization constant. We note that, contrary
to what happens for the case of a linear coupling, we obtain a
contribution whose strength is independent of $\epsilon$.  

In the case of Neumann boundary conditions:
\begin{equation}
 \tilde\mu_0(k_\parallel, \epsilon) = {\bar\mu}_0 +
 \frac{g^2}{4\pi} \frac{1}{k_\parallel}\arctan{\big(\frac{ k_\parallel}{2
m}\big)} + \frac{g^2}{4\pi} 
\sum_{n\geq 1}\left[\frac{1}{k_\parallel}\arctan{\big(\frac{\epsilon k_\parallel}{2\sqrt{n^2\pi^2 +
\epsilon^2 m^2}}\big)} - \frac{\epsilon}{2 n \pi}\right]\;, 
\end{equation}
which in the case of a massless microscopic field becomes:
\begin{equation}
 \tilde\mu_0(k_\parallel, \epsilon) = {\bar\mu}_0 \,+\, g^2 \,
 \left\{ \frac{1}{8 \, k_\parallel} + \frac{1}{4\pi} 
\sum_{n\geq 1}\Big[\frac{1}{k_\parallel}\arctan{\big(\frac{\epsilon k_\parallel}{2 n \pi}}\big) - 
\frac{\epsilon}{2 n \pi}\Big] \right\}\;. 
\end{equation}
We note that the previous series can be summed, the exact result
being
\begin{equation}
\tilde\mu_0(k_\parallel, \epsilon) = {\bar\mu}_0 - \frac{g^2}{8 \pi k_\parallel}\left\{  
\frac{\epsilon \gamma k_\parallel}{\pi} + 2 {\rm Arg}\Gamma (\frac{i k_\parallel \epsilon}{2\pi})\right\}.
\end{equation}
The computation of $\tilde\mu_2$ can be performed along similar lines, although is much more cumbersome
and we will not present the details here.

To summarize, in this Section we have shown that the nonlocal interaction
between the mirror and the scalar field admits a derivative expansion  of
the kind given in  Eq.(\ref{eq:vnldelta}), where the coefficients
$\tilde\mu_i$ depend not only on $k_{\parallel}$ but also on the width $\epsilon$
of the mirror.  It is possible to adjust the relation of the coupling
constants and $\epsilon$ in such a way that the leading term is finite in
the limit $\epsilon\rightarrow 0$, while the nonleading contributions are
suppressed by powers of $\epsilon$. Potentials proportional to the
$\delta$-function and its derivatives have been considered previously by
other authors (see for instance \cite{miltondelta,bordagdelta}). We have
shown here that these potentials arise naturally, in concrete examples, as
the leading terms in a derivative expansion of the nonlocal effective
interaction.

\section{The Casimir energy for generalized $\delta$-potentials}\label{sec:deltapot}

In this section,  we compute the Casimir energy that results from a derivative 
expansion of the nonlocal effective action, namely, when the interaction
term at $x_d=0$, after Fourier transforming the parallel coordinates, has
the local form:
\begin{equation}\label{eq:fexp}
 S_I^{(0)}(\varphi) = \frac{1}{2} \int \frac{d^dk_\parallel}{(2\pi)^d} \,
\int dx_d \,\left[\tilde\mu_0 \delta(x_d) |\widetilde{\varphi}(k_\parallel,x_d)|^2  + \tilde\mu_1 \delta'(x_d)
|\widetilde{\varphi}(k_\parallel,x_d)|^2 + \tilde\mu_2
\delta(x_d)|\widetilde{\varphi}'(k_\parallel,x_d)|^2 
\right],
\end{equation}
where $\tilde\mu_0$, $\tilde\mu_1$, $\tilde\mu_2$ are arbitrary real (in Euclidean spacetime) 
functions of $k_{\parallel}$ and $\epsilon$. 
Note that Eq.(\ref{eq:fexp}) proceeds from a nonlocal action in coordinate
space:
\begin{eqnarray}\label{eq:fexp1}
 S_I^{(0)}(\varphi) &=& \frac{1}{2} \int d^dx_\parallel\, \int d^dy_\parallel\,
\int dx_d \,\Big[\mu_0 \delta(x_d) \varphi(x_\parallel,x_d)
\varphi(y_\parallel,x_d)  \nonumber\\
& & + \mu_1 \delta'(x_d) \varphi(x_\parallel,x_d) \varphi(y_\parallel,x_d) 
+ \mu_2 \delta(x_d) \varphi'(x_\parallel,x_d) \varphi'(y_\parallel,x_d) 
\Big],
\end{eqnarray}
where now $\mu_0$, $\mu_1$, $\mu_2$ depend on $x_\parallel -y_\parallel$.
The interaction term for the remaining mirror, $S_I^{(a)}$, is obtained by
a simple shift. 

The terms containing derivatives of the $\delta$-function are expected to
be suppressed by powers of $\epsilon$, however, for the sake of generality,
we will first compute the Casimir energy for the above interaction terms
exactly, performing the expansion in powers of $\epsilon$ afterwards. One
of the reasons for this procedure is that the knowledge of the exact
Casimir energy for (\ref{eq:fexp1}) may be useful in other circumstances,
not necessarily related to the effective models we are considering here. 

Introducing the matrix 
\begin{equation}
{\bf M}=\bigg(\begin{array}{cc}
  \mu_0 & -\mu_1 \\
-\mu_1 & \mu_2
\end{array}\bigg)\, , 
 \label{matrixM}\end{equation}
(a function of $x_\parallel-y_\parallel$) we can write (\ref{eq:fexp1}) as follows:
\begin{equation}
 S_I^{(0)}(\varphi) =  \frac{1}{2} \int d^{d}x_\parallel \int d^{d}y_\parallel
\left(\varphi(x_\parallel,0), \varphi'(x_\parallel,0)\right)^T {\bf
M}(x_\parallel-y_\parallel)
\left(\varphi(y_\parallel,0), \varphi'(y_\parallel,0)\right),
\end{equation}
and a similar expression for $S_I^{(a)}$.

Then, we introduce two sets of auxiliary fields $\xi_1^{(0)}$, $\xi_2^{(0)}$ and $\xi_1^{(a)}$, $\xi_2^{(a)}$,
in order to write
 $$\exp\{-S_I(\varphi)\} = 
\frac{1}{{\cal N}}\int {\cal D}\xi_1^{(0)}{\cal D}\xi_2^{(0)}{\cal
D}\xi_1^{(a)}{\cal D}\xi_2^{(a)} $$
$$
\times  \exp\Big\{-\frac{1}{2} \int d^dx_\parallel \int d^dy_\parallel  
\big[\xi^{(0)T}(x_\parallel) {\bf M}^{-1}(x_\parallel-y_\parallel) \xi^{(0)}(y_\parallel)
+\xi^{(a)T}(x_\parallel) {\bf M}^{-1}(x_\parallel-y_\parallel)
\xi^{(a)}(y_\parallel)\big]
$$
\begin{equation}
+  i \int d^{d + 1}x\big[J^{(0)}(x) + J^{(a)}(x)\big]
\varphi(x)\Big\},
\label{x}
\end{equation}
with
\begin{eqnarray}
 J^{(0)}(x) &=& \xi_1^{(0)}(x_\parallel) \delta(x_d) - \xi_2^{(0)}(x_\parallel)\delta'(x_d)\nonumber \\
J^{(a)}(x) &=& \xi_1^{(a)}(x_\parallel) \delta(x_d - a) -
\xi_2^{(a)}(x_\parallel)\delta'(x_d - a) \;,
\end{eqnarray}
and ${\mathcal N}$ an irrelevant constant.
Note that the representation (\ref{x}) makes sense when all the eigenvalues
of ${\bf M}$ are greater than zero (they are real, since the matrix is
Hermitian).

Using Eq.(\ref{x}), we may write the generating functional as follows:
\begin{eqnarray}
 {\cal Z} &=& \int {\cal D}\varphi{\cal D}\xi \exp\left\{- S_0(\varphi) - 
\frac{1}{2} \int d^dx_\parallel\,d^dy_\parallel\left(\xi^{(0)T} {\bf M}^{-1}
\xi^{(0)} + \xi^{(a)T} {\bf M}^{-1} \xi^{(a)} \right)\right\}\nonumber \\
&\times& \exp\left\{i\int d^{d+1}x (J^{(0)} + J^{(a)})\varphi\right\},
\end{eqnarray}
where we omitted, for the sake of clarity, writing all the arguments. 

After integrating the field $\varphi$ we get
\begin{eqnarray}\frac{{\cal Z}}{{\cal Z}_0} &=& \int  {\cal D}\xi
\exp{\left\{-\frac{1}{2}\int d^dx_\parallel\,d^dy_\parallel \left[\xi^{(0)T}{\bf M}^{-1}\xi^{(0)}
+\xi^{(a)T}{\bf M}^{-1}\xi^{(a)}\right] \right\}} \nonumber \\
&\times& \exp{\left\{-\frac{1}{2}\int d^{d+1}x \int d^{d+1}y \left(J^{(0)}(x) +
J^{(a)}(x)\right) \Delta(x,y)\left(J^{(a)}(y) +
J^{(a)}(y)\right)\right\}},
\end{eqnarray}
where \begin{equation}{\cal Z}_0 = \int{\cal D}\varphi
e^{-S_0(\varphi)} ~~~~ \mbox{and} ~~~~ \Delta(x,y) = \langle x\vert
(-\partial^2)^{-1}\vert y\rangle.
\end{equation}

We first evaluate the term quadratic in the  currents more explicitly,
\begin{eqnarray}
Q &=& \frac{1}{2}\int d^{d+1}x\, d^{d+1}y\left(J^{(0)}(x) + J^{(a)}(x)\right)
\Delta(x,y)\left(J^{(0)}(y) + J^{(a)}(y)\right)  \nonumber \\
&=& \frac{1}{2}\int d^dx_\parallel d^dy_\parallel
\left[\xi_1^{(0)}(x_\parallel) \Delta(x_\parallel, 0; y_\parallel,
0) \xi_1^{(0)}(y_\parallel) + \xi_2^{(0)}(x_\parallel)
\partial_d\partial_d' \Delta(x_\parallel, 0; y_\parallel, 0)
\xi_2^{(0)}(y_\parallel)\right. \nonumber \\
&+&\xi_1^{(0)}(x_\parallel) \partial_d' \Delta(x_\parallel, 0;
y_\parallel, 0) \xi_2^{(0)}(y_\parallel) + \xi_2^{(0)}(x_\parallel)
\partial_d \Delta(x_\parallel, 0; y_\parallel, 0)
\xi_1^{(0)}(y_\parallel) \nonumber \\
&+& \xi_1^{(a)}(x_\parallel) \Delta(x_\parallel, 0; y_\parallel,
0)\xi_1^{(a)}(y_\parallel) +
\xi_2^{(a)}(x_\parallel)
\partial_d\partial_d' \Delta(x_\parallel, 0; y_\parallel,
0)\xi_2^{(a)}(y_\parallel) \nonumber \\
&+&\xi_1^{(a)}(x_\parallel) \partial_d' \Delta(x_\parallel, 0; y_\parallel,
0)\xi_2^{(a)}(y_\parallel) + \xi_2^{(a)}(x_\parallel)
\partial_d \Delta(x_\parallel, 0; y_\parallel,
0)\xi_1^{(a)}(y_\parallel) \nonumber \\
&+& \xi_1^{(0)}(x_\parallel) \Delta(x_\parallel, 0; y_\parallel,
a)\xi_1^{(a)}(y_\parallel) +
\xi_1^{(a)}(x_\parallel) \Delta(x_\parallel, a; y_\parallel,
0)\xi_1^{(0)}(y_\parallel) \nonumber \\
&+&\xi_1^{(0)}(x_\parallel) \partial_d' \Delta(x_\parallel, 0; y_\parallel,
a)\xi_2^{(a)}(y_\parallel) + \xi_2^{(0)}(x_\parallel)
\partial_d \Delta(x_\parallel, 0; y_\parallel,
a)\xi_1^{(a)}(y_\parallel) \nonumber \\
&+&\xi_1^{(a)}(x_\parallel) \partial_d' \Delta(x_\parallel, a; y_\parallel,
0)\xi_2^{(0)}(y_\parallel) + \xi_2^{(a)}(x_\parallel)
\partial_d \Delta(x_\parallel, a; y_\parallel,
0)\xi_1^{(0)}(y_\parallel) \nonumber \\
&+& \left. \xi_2^{(0)}(x_\parallel)
\partial_d\partial_d' \Delta(x_\parallel, 0; y_\parallel,
a)\xi_2^{(a)}(y_\parallel) +\xi_2^{(a)}(x_\parallel)
\partial_d\partial_d' \Delta(x_\parallel, a; y_\parallel,
0)\xi_2^{(0)}(y_\parallel) \right] \;,
\end{eqnarray}
and then, introducing Fourier transforms
in the parallel coordinates, we can show that
\begin{eqnarray}
Q &=& \frac{1}{2} \int \frac{d^dk_\parallel}{(2\pi)^d} \left[ {\tilde
\xi}_1^{(0)*} \frac{1}{2k_\parallel} {\tilde \xi}_1^{(0)} + {\tilde
\xi}_2^{(0)*} \Lambda(k_\parallel) {\tilde \xi}_2^{(0)} \right. \nonumber \\
&+& {\tilde \xi}_1^{(a)*} \frac{1}{2k_\parallel} {\tilde \xi}_1^{(a)} + {\tilde
\xi}_2^{(a)*} \Lambda(k_\parallel) {\tilde \xi}_2^{(a)} \nonumber \\
&+& {\tilde \xi}_1^{(0)*} \frac{e^{-k_\parallel a}}{2k_\parallel} {\tilde \xi}_1^{(a)}
-{\tilde
\xi}_2^{(0)*} \frac{k_\parallel}{2} e^{-k_\parallel a} {\tilde \xi}_2^{(a)} \nonumber \\
&+& {\tilde \xi}_1^{(a)*} \frac{e^{-k_\parallel a}}{2k_\parallel} {\tilde
\xi}_1^{(0)} - {\tilde \xi}_2^{(a)*} \frac{k_\parallel}{2} e^{-k_\parallel a} {\tilde
\xi}_2^{(0)}\nonumber\\
&-& {\tilde \xi}_1^{(0)*} \frac{e^{-k_\parallel a}}{2} {\tilde
\xi}_2^{(a)} +  {\tilde \xi}_2^{(0)*} \frac{e^{-k_\parallel a}}{2} {\tilde
\xi}_1^{(a)}\nonumber \\
&+& \left. {\tilde \xi}_1^{(a)*} \frac{e^{-k_\parallel a}}{2} {\tilde
\xi}_2^{(0)} -  {\tilde \xi}_2^{(a)*} \frac{e^{-k_\parallel a}}{2} {\tilde
\xi}_1^{(0)}
 \right],\label{QF}\end{eqnarray} 
where $\Lambda(k_\parallel)\equiv \frac{1}{\epsilon}-\frac{k_\parallel}{2}$. 
To compute $\Lambda(k_\parallel)$ we used that
\begin{equation}
\partial_d\partial_d' \Delta(x_\parallel, 0; y_\parallel, 0)=\delta(x_\parallel-y_\parallel)
\delta(0)-\int\frac{d^dk_\parallel}{(2\pi)^d}\frac{k_\parallel}{2}e^{ik_\parallel(x_\parallel -
y_\parallel)}\, ,
\end{equation}
and we approximated $\delta(0)\approx 1/\epsilon$. This follows 
from recalling how we obtained the derivative expansion: an approximant 
of the $\delta$ was replaced by its limit.
A Fourier transformation of the above then yields the
expression used for $\Lambda(k_\parallel)$. 

On the other hand, the first derivatives of the free propagator are  
ill-defined at $x_d =x'_d$.
In order to obtain Eq.(\ref{QF}), we have used a symmetric limit regularization so
that $\partial_d\Delta(0;0)=\partial'_d\Delta(0;0)=0$.

We can now compute the vacuum energy ${\cal E}_0$ as follows,
\begin{equation}
{\cal E}_0 = - \lim_{T, L \rightarrow \infty} \frac{1}{TL^d}
\ln\left(\frac{{\cal Z}}{{\cal Z}_0} \right) =
\frac{1}{2}\int\frac{d^dk_\parallel}{(2\pi)^d}\ln{\det}({\bf A}(k_\parallel))\,\, ,
\end{equation}
where
\begin{eqnarray}
{\bf A}(k_\parallel)&=&\left(\begin{array}{cccc}
  ({\bf \tilde M}^{-1})_{11} + \frac{1}{2k_\parallel} & ({\bf \tilde M}^{-1})_{12}  & \frac{e^{-k_\parallel a}}{2k_\parallel} & -\frac{e^{-k_\parallel a}}{2} \\
({\bf \tilde M}^{-1})_{21}  & ({\bf \tilde M}^{-1})_{22} +\Lambda(k_\parallel) & \frac{e^{-k_\parallel a}}{2} & -\frac{k_\parallel}{2} e^{-k_\parallel a} \\
\frac{e^{-k_\parallel a}}{2k_\parallel} & \frac{e^{-k_\parallel a}}{2} &({\bf \tilde M}^{-1})_{11} + \frac{1}{2k_\parallel} &({\bf\tilde
M}^{-1})_{12} \\
-\frac{e^{-k_\parallel a}}{2} 
& -\frac{k_\parallel}{2} e^{-k_\parallel a} & ({\bf \tilde M}^{-1})_{12} &
({\bf \tilde M}^{-1})_{22} +\Lambda(k_\parallel)
\end{array}\right)\nonumber\\
&\equiv & 
\Bigg(\begin{array}{cc}
  {\bf A}_{11}  & e^{-k_\parallel a}{\bf A}_{12} \\
e^{-k_\parallel a}{\bf A}_{21}  & {\bf A}_{22} 
\end{array}\Bigg)\; ,
\end{eqnarray}
where ${\bf A}_{ij}$ are four $2\times 2$ block-matrices and ${\bf\tilde M}$ is the Fourier transform of 
the matrix  ${\bf M}$ defined in Eq.(\ref{matrixM}).

Finally, the subtracted (i.e., without self-energies) energy can be written as
\begin{equation}\label{eq:energydelta}
{\tilde {\cal E}}_0 = \frac{1}{2} \int \frac{d^dk_\parallel}{(2\pi)^d}
\ln{\det} \left[I - e^{-2k_\parallel a}{\bf \Gamma}(k_\parallel) \right],
\end{equation}
where
\begin{equation}
{\bf \Gamma}(k_\parallel) = ({\bf A}_{11})^{-1} {\bf A}_{12}({\bf
A}_{22})^{-1} {\bf A}_{21},
\end{equation}
and
\begin{equation}
{\bf A}_{11}={\bf A}_{22}=\Bigg(\begin{array}{cc}
  ({\bf \tilde M}^{-1})_{11} + \frac{1}{2k_\parallel} & ({\bf \tilde M}^{-1})_{12}  \\
({\bf \tilde M}^{-1})_{21} & ({\bf \tilde M}^{-1})_{22} +\Lambda(k_\parallel)
\end{array}\Bigg) ,
\end{equation}

\begin{equation}
{\bf A}_{12} = {\bf A}_{21}^T =  \frac{1}{2}  \Bigg(\begin{array}{cc}
  \frac{1}{k_\parallel} & - 1 \\
 1 & -k_\parallel
\end{array}\Bigg) .
\end{equation}

Eq.(\ref{eq:energydelta}) is the main result in this section; it gives the exact Casimir energy for 
the generalized $\delta$-potentials, as a function of the coefficients that
determine the effective interaction term. It is interesting to emphasize that the structure of the vacuum energy
in Eq.(\ref{eq:energydelta}) is similar to the Lifshitz formula for the electromagnetic field in the presence of
anisotropic materials, in which it is necessary to introduce $2\times 2$ reflection matrices to take
into account the mixing between TE and TM modes \cite{dalvit08}. Here the $2\times 2$ matrices come from the introduction of
two auxiliary fields on each mirror to describe the effective action (see Eq.(\ref{x})). 

The presence of a divergence, when $\epsilon \to 0$, in
$\Lambda(k_\parallel)$ does not introduce any divergence in the Casimir
energy. Indeed, the inverses of the matrices 
${\bf A}_{11}$ and ${\bf A}_{22}$ are finite in that limit $\epsilon\rightarrow 0$.
Computing explicitly the determinant in Eq.({\ref{eq:energydelta}}), the final
expression becomes
\begin{equation}
{\tilde {\cal E}}_0 = \frac{1}{2} \int \frac{d^dk_\parallel}{(2\pi)^d}
\ln \left[1- e^{-2ak_\parallel}F(k_\parallel,\epsilon)-  e^{-4ak_\parallel}G(k_\parallel,\epsilon) \right], 
\label{energy general}
\end{equation}
where $F$ and $G$ depend on $k_\parallel$ and $\epsilon$ explicitly 
and also implicitly through the coefficients $\tilde\mu_i$.
Although it is possible to obtain general expressions 
for $F$ and $G$, the result is a rather lengthy expression, which is not very illuminating. 
Thus we will analyze two interesting particular cases. 

Let us first consider the calculation of the Casimir energy in the
framework of the derivative expansion introduced in the previous section.
As already mentioned,  on general grounds we expect  the terms containing
higher derivatives to be suppressed by powers of $\epsilon$. To make this
point explicit, we write
$\tilde\mu_0=\lambda_0, \tilde\mu_1=\epsilon\lambda_1$ and $\tilde\mu_2=\epsilon^2\lambda_2$,
where $\lambda_i$ are of the same order of magnitude in the limit $\epsilon\rightarrow 0$.  We then evaluate
the determinat in Eq.({\ref{eq:energydelta}})
exactly, and expand the result in powers of $\epsilon$. After a long but
nevertheless straightforward calculation, the expansion for the Casimir
energy adopts the form
\begin{equation}
{\tilde {\cal E}}_0 = \frac{1}{2} \int \frac{d^dk_\parallel}{(2\pi)^d}
\ln \left[1- e^{-2ak_\parallel}(f_0+\epsilon f_1 +\epsilon^2 f_2 +O(\epsilon^3))\right], 
\label{expansionenergy}
\end{equation}
where
\begin{eqnarray}
f_0&=&\frac{\lambda_0^2}{(2k_\parallel+\lambda_0)^2}\nonumber\\
f_1&=&-4\lambda_0\lambda_1^2\frac{k_\parallel}{(2k_\parallel+\lambda_0)^3}\nonumber\\
f_2&=&\frac{4\lambda_1^2k_\parallel}{(2k+\lambda_0)^4}\left[\lambda_0(\lambda_0\lambda_2-\lambda_1^2)
+k_\parallel(2\lambda_0\lambda_2+\frac{\lambda_0^2}{2}+\lambda_1^2)
+k^2_\parallel\lambda_0\right]\, .
\label{expansion2}
\end{eqnarray}
The coefficients $\lambda_i$, depending on $k_\parallel$,  encode all
the information about the interaction between  the vacuum field and the
microscopic degrees of freedom living on the mirrors. The outcome is
dominated by the usual $\delta$-potential result, but with a
$k_\parallel$-dependent strength. It is worth noting that, up to this order,
there are no terms proportional to $e^{-4ak_\parallel}$  in the Casimir energy.
These terms do appear in the fourth order.

In order to illustrate the kind of
corrections induced by the finite width of the mirrors, let us assume that the
coefficients $\lambda_i$ are approximately constants. Then the Casimir energy
for $d=3$ can be written, to first order in $\epsilon$ as follows:
\begin{equation}
{\tilde {\cal E}}_0 = -\frac{1}{a^3}\left[I_1(\lambda_0 a)+\frac{\epsilon}{a}(\frac{\lambda_1}{\lambda_0})^2 I_2(\lambda_0 a)\right ]+O(\epsilon ^2)\, ,
\end{equation}
where we introduced the coefficient functions $I_i$, which can be easily
evaluated numerically. They are plotted in Figs.1 and 2. Both of them are monotonic 
and positive definite functions,  interpolating between $0$ and a finite value for $\lambda_0 a
\rightarrow\infty$. The leading term reproduces the Casimir result for perfectly conducting
mirrors in this limit, while the second one introduces a correction that
falls off faster (with an extra power of the distance).

\begin{figure}[h!t]
\centering
\includegraphics[width=9.5cm]{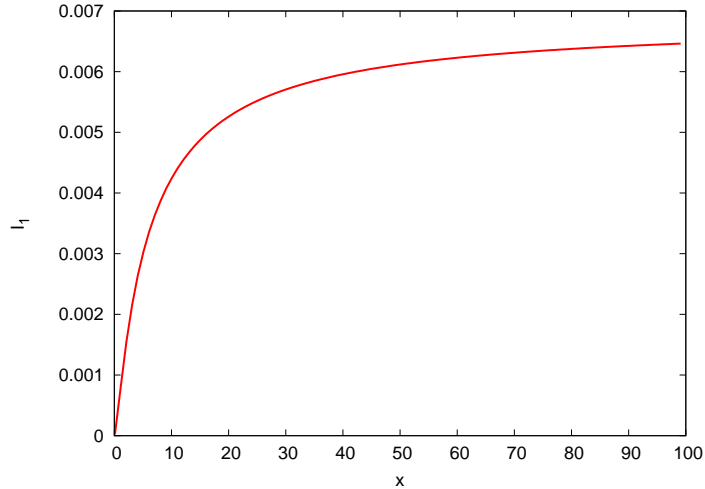}
\caption{Coefficient function $I_1$ as a function of $x = {\lambda}_0 a$. This  
function reproduces the Casimir result for perfect conducting plates in the limit $x\rightarrow \infty$, 
$I_1(x) \rightarrow \pi^2/1440$. }
\label{fig1}
\end{figure}

\begin{figure}[h!t]
\centering
\includegraphics[width=9.5cm]{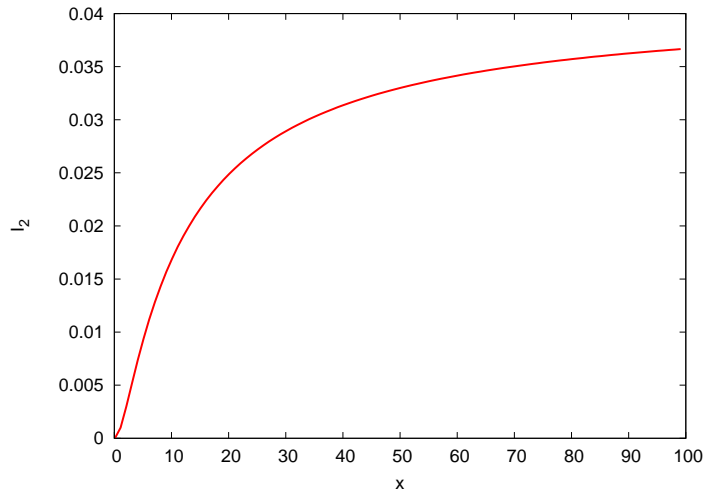}
\caption{Coefficient function $I_2$ as a function of $x = {\lambda}_0 a$. This function 
approaches $0.0366$ in the limit $x\rightarrow \infty$.}
\label{fig2}
\end{figure}

As a second example, we now consider the vacuum energy for the generalized
$\delta$-potentials, assuming that the coefficients $\tilde\mu_i$ do not depend
on  $\epsilon$. In this situation
we obtain
\begin{equation}
{\tilde {\cal E}}_0 = \frac{1}{2} \int \frac{d^dk_\parallel}{(2\pi)^d}
\ln \left[1-\frac{(\tilde\mu_0\tilde\mu_2-\tilde\mu_1^2)^2 \, e^{-2ak_\parallel}}{[(2k_\parallel+\tilde\mu_0)\tilde\mu_2-
\tilde\mu_1^2]^2} \right]= \frac{1}{2} \int \frac{d^dk_\parallel}{(2\pi)^d}
\ln \left[1-\frac{\tilde\mu_{eff}^2 \, e^{-2ak_\parallel}}{[2k_\parallel+\tilde\mu_{eff}]^2} \right],
\label{exactE}
\end{equation}
where $\tilde\mu_{eff}=(\tilde\mu_0\tilde\mu_2-\tilde\mu_1^2)/\tilde\mu_2$. 
It is interesting to note that the Casimir energy is well defined only when
the parameters $\mu_i$ are such that the coefficient that multiplies $e^{-2
a k_\parallel}$ is less than 1, and  this is the case if $\tilde\mu_{eff}>0$.  This  
condition have been found before, albeit in a different
looking but equivalent form; indeed, we have seen that, for the auxiliary field 
representation (\ref{x}) to make
sense, positivity of the eigenvalues of the matrix ${\bf M}$ is a necessary
(and sufficient) condition. 
From a physical point of view, this condition can be interpreted as
follows: the interaction term at each mirror may be diagonalized, to look
like the sum of two decoupled quadratic interaction terms, each one involving a
mixture of the field and its normal derivative at the mirror. For the
vacuum to be stable, one must have therefore non-negative eigenvalues,
since they are the coefficients that affect each decoupled term.
Vanishing eigenvalues, on the other hand, are not forbidden physically, 
rather, one should represent them with just one auxiliary field. Otherwise
the redundancy pops up in the form of a zero mode.
 
It is remarkable that Eq.(\ref{exactE}) corresponds to the Casimir energy
for a usual  $\delta$-potential (i.e. without derivatives of the $\delta$-function)
with an effective coefficient  given  by $\tilde\mu_{eff}$. However, it is also worth stressing that  
this equation has been derived assuming
a particular regularization for $\Lambda(k_{\parallel})$. While this
regularization is well justified in the case of the derivative expansion, a
formal calculation which started from the generalized $\delta$-potentials,
could give different, regularization dependent results, without any
immediate physical reason to chose one from another. 
For example, in the framework of dimensional
regularization one would obtain $\Lambda(k_\parallel)= \alpha-\frac{k}{2}$,
with $\alpha$ an arbitrary constant. 

Different regularizations of this object correspond, physically,
to imposing different boundary conditions for the propagator of the vacuum
field at the mirror. If the concrete model for the finite width mirror is
unknown, this lack of information manifests itself in the fact that one has
many regularizations available, and they give rise to different values of
the
energy. However, one knows that, what makes sense physically is not
the regularization used; rather, it is the boundary condition it produces
on the propagator. Then, one may regard the boundary condition for the
propagator as a renormalization condition which hides the ignorance on the
details of the model into a bare coefficient function \cite {prep}. 

\section{Discussion}\label{sec:disc}
We have shown, in concrete examples, how the nonlocal induced action which
results from the integration of the microscopic fields (that represent the
media composing the mirrors) can be expanded to produce a local action;
i.e., one that has point-like support. This means that it may be written as
terms involving the $\delta$-function and its derivatives. 

Equipped with the general form of that local action, we then derived the
Casimir energy for the vacuum scalar field. A conceptually interesting point is that the presence of 
derivatives of the $\delta$-function, in the effective action,
produces a final result for the Casimir energy that can be written in the form
of the Lifshitz formula for the electromagnetic field with $2\times 2$ reflection matrices. This 
analogy is a biproduct of the representation of the effective action in
terms of two auxiliary fields 

When the coefficients from that local action come from a
microscopic model, we have shown that the result may be consistently 
expanded in powers of the width of the mirrors, producing a result which
may be interpreted as a Dirichlet-like energy plus sub-leading corrections. 
It would be interesting to generalize these results to the realistic case of the 
electromagnetic field coupled to Dirac fields describing charges on the mirrors.

Besides, we considered also the case when the coefficients are assumed to
be independent of the width of the mirrors.  In this case,
the exact result adopts a quite simple form: the vacuum energy coincides with the one
given by the usual $\delta$-potential, with an effective coupling. That is to say, the 
effect of the terms proportional to derivatives of the $\delta$-function in the effective action
is to renormalize the coupling of the usual $\delta$ potential. The last
result depends, in principle, on the particular regularization used to
handle these (highly singular) potentials. However, when one abandons the
description in terms of the coefficients for those terms, in favour of
another in terms of the boundary conditions on the propagator, the apparent ambiguity disappears \cite{prep}.

\section*{Acknowledgements}
C.D.F. thanks CONICET, ANPCyT and UNCuyo for financial support. The work of
F.D.M. and F.C.L was supported by UBA, CONICET and ANPCyT.

\end{document}